\documentclass[11pt]{cernrep_a}
\usepackage{graphicx}
\usepackage{floatflt}

\newcommand{\Et}{{E_t}}
\newcommand{\Pt}{{P_t}}

\newcommand{\gpj}{~``$\gamma+jet$''~}
\newcommand{\pth}{\hat{p}_{\perp}^{\;min}}
\newcommand{\hn}{\hspace*{-5mm}}
\newcommand{\hnn}{\hspace*{3.2mm}}
\newcommand{\Ptg}{\Pt^{\gamma}}
\newcommand{\lt}{\!<\!}
\newcommand{\gt}{\!>\!}

\suppressfloats[!]

\def\baselinestretch{1.0}

\begin{document}
\thispagestyle{empty}
 
\vskip-5mm
 
\begin{center}
{\Large JOINT INSTITUTE FOR NUCLEAR RESEARCH}
\end{center}
 
\vskip10mm
 
\begin{flushright}
JINR Communication  \\
E1-2001-261 \\
hep-ex/0108050
\end{flushright}
 
\vspace*{3cm}
 
\begin{center}
\noindent
{\Large{
\bfseries On the possibility of discrimination between
$\pi^0, \eta, \omega, K^0_s$ mesons and a photon  based on
 the calorimeter information in the CMS detector}}\\[10mm]
{\large D.V.~Bandourin$^{1\,\dag}$, V.F.~Konoplyanikov$^{2\,\ast}$,
N.B.~Skachkov$^{3\,\dag}$}  
 
\vskip 0mm
 
{\small
{\it
E-mail: (1) dmv@cv.jinr.ru, (2) kon@cv.jinr.ru, (3) skachkov@cv.jinr.ru}}\\[3mm]
$\dag$ \large \it Laboratory of Nuclear Problems \\
\hspace*{-5mm} $\ast$ \large \it Laboratory of Particle Physics        
\end{center}
 
\vskip 14mm
\begin{center}
\begin{minipage}{150mm}
\centerline{\bf Abstract} 
~\\[-1pt]
\noindent
The possibility of separating the $\pi^0, \eta, \omega$ and  $K^0_s$ meson background
from the signal photons produced directly in $pp$ collisions is
analyzed. The rejection factors for two 
pseudorapidity regions $0.4\lt\eta\lt1.0$ and $1.6\lt\eta\lt2.0$ and
six $\Et$ values, $20, 40, 60, 80, 100$ and $200~GeV$, are calculated
for a case when only the calorimeter information is used.
The cases of $\eta, \omega$ and $K^0_s$ mesons decaying through the neutral and
charged channels are considered separately.
\end{minipage}
\end{center}       

\newpage
 
\setcounter{page}{1}   
%
\section{Introduction.}
%
\noindent

This work is a continuation of our previous publications on the study of \gpj events
at LHC energies \cite{RDMS1} -- \cite{GLU}. 
In those papers it was shown that a process of direct photon production (with $\Ptg\gt40 ~GeV/c$) in 
association with one jet, caused mainly by the Compton-like ~$q g\to \gamma +q$~ and annihilation 
 ~$q \bar{q}\to \gamma + g$ subprocesses, has a considerable background due to other QCD
processes that contain high $\Pt$ final state photons originating from decays of 
$\pi^0, \eta, K^0_s$ and $\omega$ mesons (see \cite{PREP5}).
The former subprocess, as was pointed out in \cite{Aur}, can be used at the LHC
for studying the gluon density in a proton in the reaction of inclusive direct photon
production and in the \gpj events (\cite{GLU}, \cite{MD1}, \cite{DMS}).
As for the decay channels of the abovementioned mesons
(see Table 1 with the PDG  data \cite{PDG} of branching ratios),
we shall consider first the neutral decay channels
($\pi^0\to2\gamma$; $K^0_s\to2\pi^0;\; \eta\to3\pi^0$ and $2\gamma$) as the next most important
background source (especially with increasing signal photon energy) after the processes with
a hard radiation of photons from quarks, i.e. bremsstrahlung photons (see \cite{PREP5}).
\\[-20pt]
\begin{table}[h]
\begin{center}
\normalsize{Table 1: Decay modes of $\pi^0, \eta, K^0_s$ and  $\omega$ mesons.}
\vskip0.2cm
\begin{tabular}{|c|c|c|}                  \hline
Particle & Br.($\%$)& Decay mode  \\\hline
$\pi^0$ & 98.8 & $\gamma\gamma$    \\
        &  1.2 & $\gamma e^+e^-$    \\\cline{1-3}
$\eta$  & 39.3 & $\gamma\gamma$    \\
        & 32.2 & $\pi^0\pi^0\pi^0$ \\
        & 23.0 & $\pi^+\pi^-\pi^0$ \\
        & 4.8  & $\pi^+\pi^-\gamma$ \\\cline{1-3}
$K^0_s$ & 68.6 & $\pi^+\pi^-$ \\
        & 31.4 & $\pi^0\pi^0$ \\\cline{1-3}
$\omega$& 88.8 & $\pi^+\pi^-\pi^0$ \\
        &  8.5 & $\pi^0\gamma$ \\\cline{1-3}
\end{tabular}
\end{center}
\end{table}
~\\[-15pt]
We shall study a question of what level of accuracy can be achieved in suppression
of the contribution from these neutral decay channels if we confine ourselves
only to the information from the CMS electromagnetic calorimeter (ECAL).
Possible use of preshower detector information is not considered here as
it was a subject of another publication \cite{BarANN}.

Another group of decay channels that contain charged pions in the final state is
less difficult to be suppressed. As will be shown below, their contribution to the background
can be discriminated with a good efficiency on the basis of the
hadronic calorimeter (HCAL) data.

The results presented  here are obtained from the simulation with the GEANT based package CMSIM
\footnote{The results in Sections 2 and 3 of this paper are obtained with CMSIM
versions 111 and 116 respectively.}.
We carried out a few simulation runs including\\
(1) four particle types: single photons $\gamma$ and
$\pi^0, \eta, K^0_s$ mesons; \\
(2) six $\Et$ values for each of these particles: $20, 40, 60,$ $80, 100$ and $200 ~GeV$;\\
(3) two pseudorapidity regions: $0.4\!<\!\eta\!<\!1.0$ (Barrel) and $1.6\!<\!\eta\!<\!2.0$ (Endcap).\\
About 4000 -- 5000 single-particle events were generated for each
$\Et$, $\eta$ interval and for each type of particle.

%
\section{Neutral decay channels.}
%

To separate the single photon events in the ECAL from the events with
photons produced in the neutral decay channels of $\pi^0,
\eta, K^0_s$ and $\omega$ mesons, two variables which
characterize the spatial distribution of $\Et$ deposition in both
cases were used:\\[5pt] 
\noindent 
\hspace*{5mm} 1. ~$\Et$
deposited inside the most energetic crystal cell in the ECAL
$\Et^{cell}_{max}$; \\ 
\hspace*{5mm} 2. The quantity $D_w$ that characterizes  spatial $\Et$ distribution 
inside the $5\times5$ crystal window around the most energetic crystal cell (see below).

The showers produced in ECAL crystal cells by the single 
photon are concentrated in a $5\times$5 array of crystals centered on the crystal with the maximum
signal (see \cite{CMS_EC}). Another picture is expected for the multiphoton
final state arising from the meson decay through the intermediate
$\pi^0$ states that would cause different  spatially  separated
centers of final state photon production. In this case the spatial 
distribution of $\Et$ deposited in such a
shower is also more likely to be different from the one produced by a single photon.
%
%

Let us define the coordinates of the ECAL shower center of gravity ($gc$) in the $\eta-\phi$ space
according to the formula:\\[-10pt]
\begin{equation}
\eta_{gc}={\left(\sum\limits_{i=1}^{N\times N} \eta_i \Et^i\right)}/\left({\sum\limits_{i=1}^{N\times N}\Et^i}\right);
\quad
\phi_{gc}={\left(\sum\limits_{i=1}^{N\times N} \phi_i \Et^i\right)}/\left({\sum\limits_{i=1}^{N\times N}\Et^i}\right)
\label{eq:COG}
\end{equation}
The sum in formula (\ref{eq:COG}) runs over ECAL crystal cells
\footnote{(with a cell size of 0.0175$\times$0.0175)}
forming the $N\times N$ window that contain a shower with
the most energetic cell in the center. Here
$\Et^i$ is $\Et$ deposited inside the $i$th crystal cell belonging to the shower.

\begin{center}
  \begin{figure}[h]
\vskip-17mm
\hspace*{-5mm} \includegraphics[width=14cm,height=11cm,angle=0]{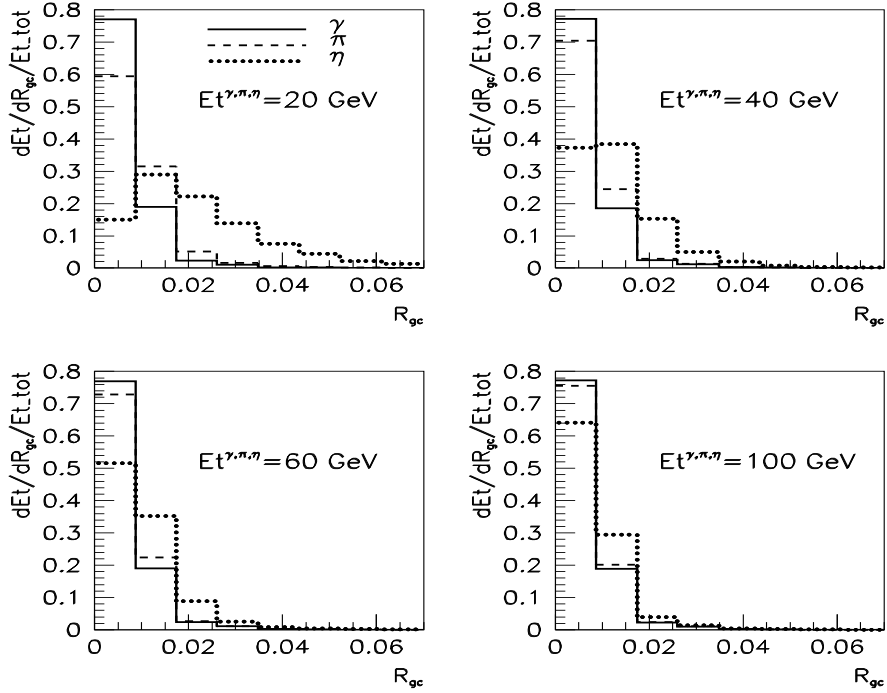}
\vskip-9mm
    \caption{The normalized $dEt/dR_{gc}/Et_{-}tot$  distribution over $R_{gc}$ of the showers produced in ECAL
by $\pi^0, \eta$ mesons and single photons $\gamma$ having $\Et=20, 40, 60$ and $100~GeV$.}
    \label{fig:etr}
  \end{figure}
\end{center}

The distributions of $\Et$ deposited in ECAL by the single photon, $\eta$ and $\pi^0$ mesons 
with initial $\Et=20, 40, 60$ and 100$~GeV$ are shown in Fig.~\ref{fig:etr}
as a function of the distance counted from the center of gravity $R_{gc}$.
One can see from Fig.~\ref{fig:etr}  that about 96$\%$ of the total photon $\Et$ deposited in ECAL, 
i.e. $Et_{-}tot$,
fits into the radius $R_{gc}=0.02$ and practically all its energy is contained inside $R_{gc}=0.045$.
From the same pictures we can conclude that for all mesons
 the total $\Et$ of showers is mainly contained inside 
the 5$\times$5 crystal cell window having the size of one 
CMS HCAL tower. So, in formula (\ref{eq:COG}) we can put with a good accuracy $N=5$.
We also observe  from this figure that the difference in spectra for 
single photons and pions decreases with growing $\Et$ and practically disappears at
$\Et=100~ GeV$, while the difference of the distributions caused by $\eta$ mesons from those for 
single photons  can still be seen at $\Et=60$ and $100 ~GeV$.
\vskip-11mm

For our further needs we introduce another useful variable.
Let us consider the distance $r_i$ of the $i$th cell from the center of gravity
of the electromagnetic shower in  ECAL, i.e.\\[-12pt]
\begin{equation}
r_i = ((\phi_i-\phi_{gc})^2+(\eta_i-\eta_{gc})^2)^{1/2},
\label{eq:ri}
\end{equation}
where $(\eta_i, \phi_i)$ are coordinates of the center of the ECAL crystal cell.
The distances $r_i$ of each crystal cell can be used to introduce a new useful quantity
(that effectively takes into account the contribution of energetic cells far from 
the center of gravity):\\[-5pt]
\begin{equation}
D_w={\left(\sum\limits_{i=1}^{25} r_i \Et^i\right)}/\left({\sum\limits_{i=1}^{25}\Et^i}\right).
\label{eq:RW}
\end{equation}

\begin{center}
  \begin{figure}[h]
\vskip-10mm
\hspace*{-7mm} \includegraphics[width=14.5cm,height=57mm,angle=0]{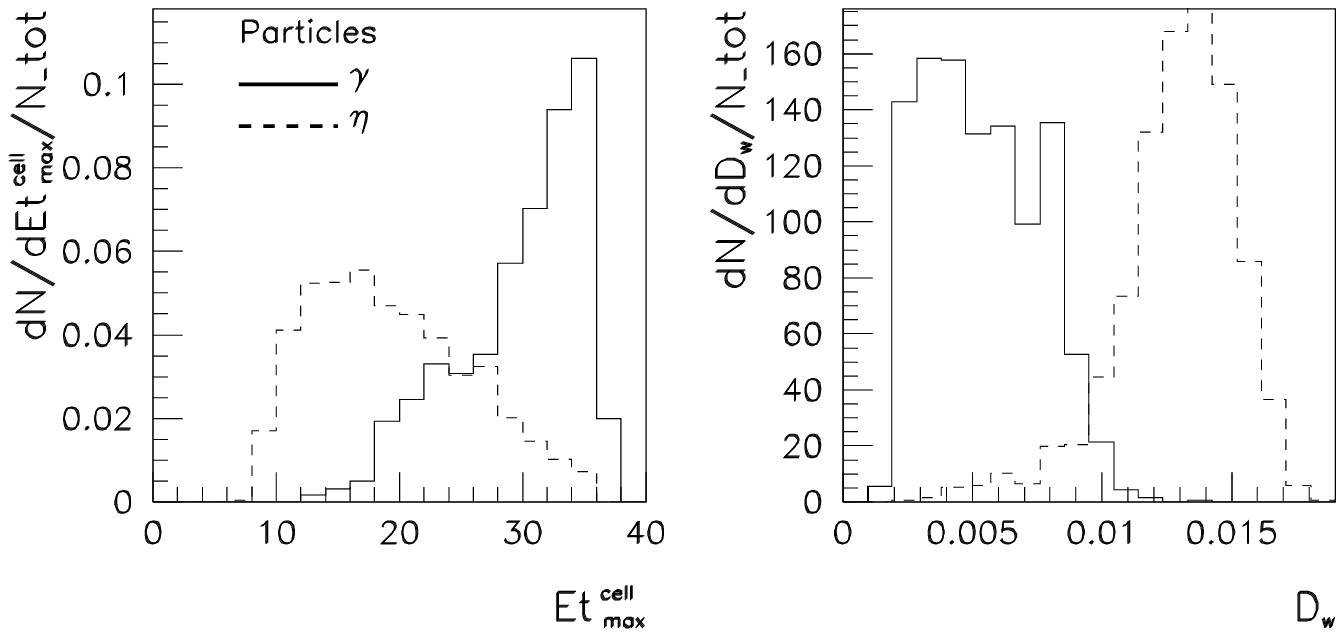}
\vskip-8mm
\caption{The normalized distributions of the number of events over $D_w$ and $\Et^{cell}_{max}$
for the single photons ($\gamma$) and for ECAL showers originating from $\eta$ mesons ($\eta$)
having $\Et=40~ GeV$.}
    \label{fig:rc_eta}
\vskip-7mm
\hspace*{-7mm} \includegraphics[width=14.5cm,height=57mm,angle=0]{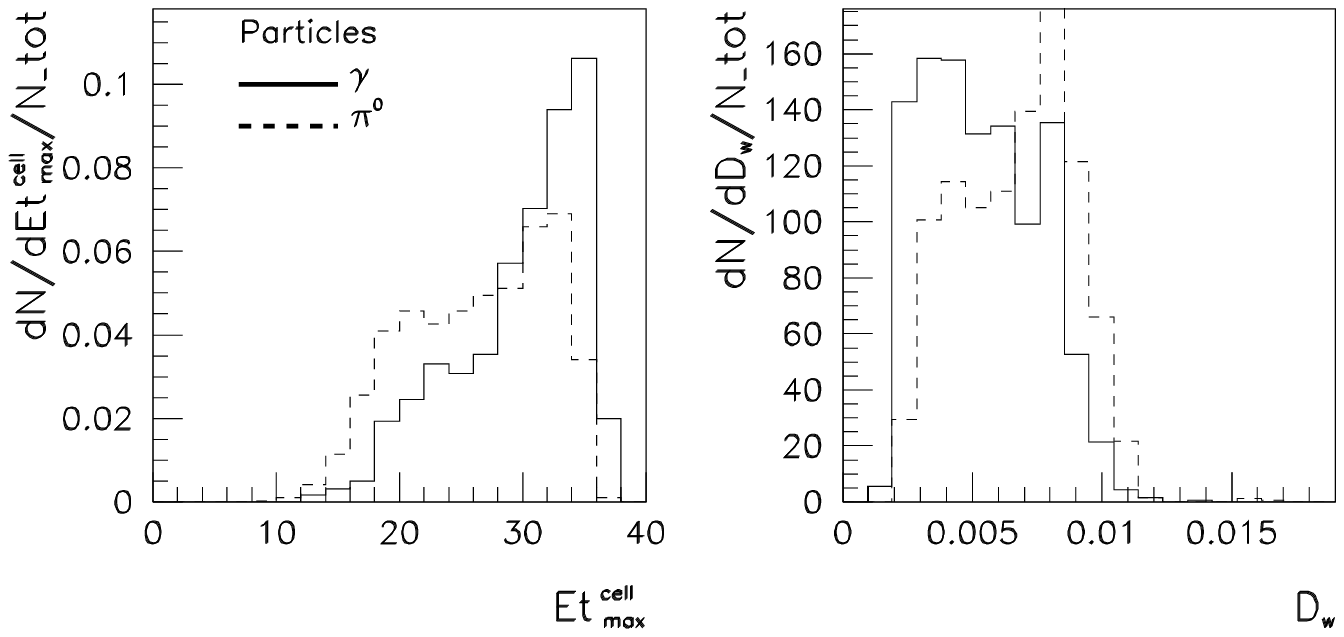}
\vskip-8mm
    \caption{The normalized distributions of the number of events over $D_w$ and $\Et^{cell}_{max}$
for the single photons ($\gamma$) and for $\pi^0$ mesons ($\pi^0$) having $\Et=40~ GeV$.}
    \label{fig:rc_pi0}
\vskip-11mm
  \end{figure}
\end{center}

For illustration,  we present  the results of CMSIM simulation with $\Et^{\gamma/mes}=40 ~GeV$ 
($mes=\pi^0, \eta$) in Figs.~\ref{fig:rc_eta} and \ref{fig:rc_pi0}. These figures contain the 
normalized distributions of $\Et^{cell}_{max}$ and $D_w$ that characterize the showers
produced by the products of $\eta$ meson (Fig.~\ref{fig:rc_eta}) and $\pi^0$ meson 
(Fig.~\ref{fig:rc_pi0}) decays. The analogous $\Et^{cell}_{max}$ and $D_w$ distributions 
in the showers produced by a single photon are also shown there for comparison.
\\[-11mm]

One can easily see that the $\eta$ meson shower spectrum over $D_w$ is strongly different from the one
of the photon. The range of $\Et^{cell}_{max}$ values where the $\eta$ meson background makes a small
contribution ($\Et^{cell}_{max}\geq 28 ~GeV$) is also clearly seen in Fig.~\ref{fig:rc_eta}.
From the result of the simulation we conclude that $90\%$ of the signal photon events at 
$\Et^{\gamma/mes}=40 ~GeV$ are concentrated in the regions 
$D_w\leq 0.0088$ and $\Et^{cell}_{max}\geq 23 ~GeV$.

So, the $\Et^{cell}_{max}$ and $D_w$ spectra of the $\eta$ meson can be effectively used for separation
of the background and the signal events because the region of their overlapping becomes much smaller
as compared with Fig.~1.

We refer to the way of separation of  signal and background events based on the differences between
distributions over the 1st variable $\Et^{cell}_{max}$ as the ``$Et_-max$'' criterion and over 
the 2nd variable $D_w$ as the ``$D_-w$'' criterion.

The situation is much worse in the case of $\pi^0$ showers as is seen in Fig.~\ref{fig:rc_pi0} where
the spectra of $\pi^0$ practically overlap with the spectra of the photon.

The neutral pion rejection efficiencies $R^{\pi^0}_{eff}$ (relative to direct photons)
\footnote{The rejection efficiency is defined as a ratio of the number of background events 
discarded by the cut, taken for a given value of signal selection efficiency $\epsilon^{\gamma}_{eff}$,
to the total number of background events.}
obtained on the basis of the $D_-w$ criterion
for different $\Et^{\gamma/\pi^0}$
are given in Table~\ref{tab:rej_pi} as a function of the chosen single
photon selection efficiencies (in $\%$)  $\epsilon^{\gamma}_{eff}$= 80, 85, 88 and 90 $\%$.
The second line in this table
``$\Et^{\gamma/\pi^0}=40 ~GeV$'' corresponds to the plots in
Fig.~\ref{fig:rc_pi0}. We see from this table that $R^{\pi^0}_{eff}$
grows by almost two-fold as $\epsilon^{\gamma}_{eff}$ decreases from 90$\%$ to 80$\%$.\\[-25pt]
\setcounter{table}{1}
\def\baselinestretch{0.90}
\begin{table}[h]
\begin{center}
\caption{Neutral pion rejection efficiencies $R^{\pi^0}_{eff}$
($\%$) obtained from CMSIM simulation by application of $D_-w$ criterion
 for five $\Et$ values of single $\gamma$ and $\pi^0$ and for different 
values of single $\gamma$ selection efficiencies $\epsilon^{\gamma}_{eff}=80 - 90\%$.
~~ $0.4<\eta^{\gamma,\pi}<1.0$.}
\label{tab:rej_pi}
\vskip0.2cm
\begin{tabular}{|lc||c|c|c|c|}   \hline
 &&\multicolumn{3}{r}{ \hspace{0.64cm} photon selection efficiencies 
$\epsilon^{\gamma}_{eff}$ } & \\\cline{3-6}
 & $\!\!\!\Et^{\gamma/\pi^0}~(GeV)\!$ &\hnn 80\%\hnn  &\hnn 85\% \hnn &\hn 88\%\hn &\hnn 90\% \hnn 
\\\hline\hline
 &$\!\!\!$ 20$\!$  & 71 & 64 & 59& 54     \\\hline
 &$\!\!\!$ 40$\!$  & 38 & 32 & 25& 22     \\\hline
 &$\!\!\!$ 60$\!$  & 33 & 26 & 20& 17     \\\hline
 &$\!\!\!$ 80$\!$  & 29 & 21 & 17& 14     \\\hline
 &$\!\!\!$ 100$\!$ & 25 & 20 & 16& 13     \\\hline
\end{tabular}
\end{center}
\end{table}
\def\baselinestretch{1.0}
~\\[-12mm]

The rejection percentages of events with
$\pi^0$ for each of the criteria $Et_-max$ and $D_-w$ are given
in Fig.~4 for different meson $\Et$
\footnote{The rejection factors for the Endcap are found to be in agreement
with those of \cite{BarEnd} (without preshower) if one 
takes into account the fact that in  Fig.~4 the rejection powers are averaged over the entire
$1.6\!~<\!\eta\!<\!2.0$ range.}.

Analogous rejection curves obtained for the $\eta$ and
$K^0_s$ meson neutral channels (for the fixed value of single
photon selection efficiency $\epsilon^{\gamma}_{eff}=90\%$) are
presented in Fig.~5 ($Et_-max$ criterion) and  in Fig.~6 ($D_-w$ criterion) for 
the interval $20 \leq \Et\leq 200~ GeV$.

By comparing Fig.~5 for the $\eta$ meson and Fig.~4 we see that for 
the same $\epsilon^{\gamma}_{eff} (=90\%)$ and the same rejection criterion $Et_-max$
~the $\pi^0$ rejection efficiency $R^{\pi^0}_{eff}$ becomes less than $30\%$  at $\Et=30 ~GeV$ while 
the $\eta$ meson rejection efficiency drops to the same level  
of $30\%$ only for $\Et\geq 90 ~GeV$.
It is also seen that \\[-10pt]
\begin{center}
\begin{figure}[htbp]
\vskip-10mm
$D_-w$ criterion works a bit better than $Et_-max$ one for the Barrel region
($0.4<\eta<1.0$).\\
\hspace*{9mm} In spite of a difficulty of neutral pion background separation from single
photon signal it is worth reminding that, as it was already mentioned in \cite{PREP5},
the $\pi^0$ background events contribution (as well as  the contribution
from $\eta, \omega$ and $K^0_s$ events),  left after 
\vskip-10mm
\hspace*{-3mm}\includegraphics[width=15cm,height=61mm,angle=0]{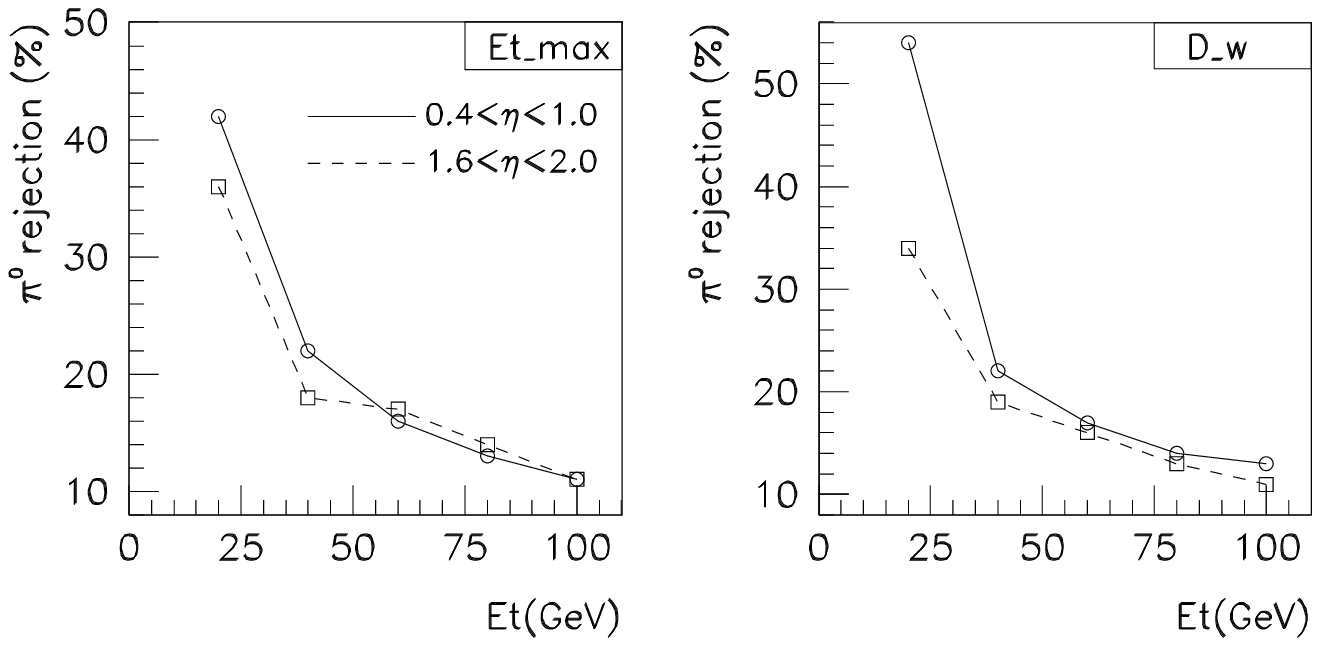}
\vskip-4mm
\hspace*{15mm} \footnotesize{Fig.~4: $\pi^0$ rejection efficiencies for $Et_-max$ and $D_-w$ criteria ($\epsilon^{\gamma}_{eff}=90\%$).}
 \label{fig:rej_pi}
%
\vskip-5mm
\hspace*{-3mm} \includegraphics[width=15cm,height=61mm,angle=0]{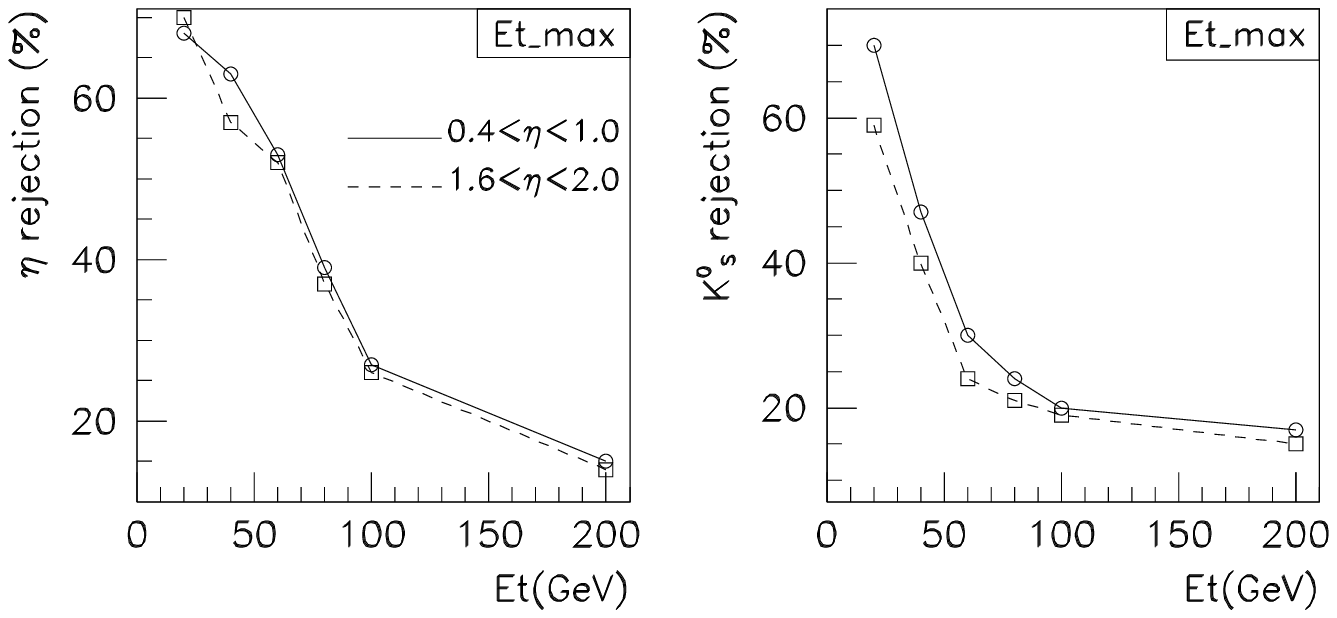}
\vskip-4mm
\hspace*{0mm} \footnotesize{Fig.~5: $\eta$ and $K^0_s$ rejection efficiencies for $\epsilon^{\gamma}_{eff}=90\%$.
Neutral decay channels only. $Et_-max$ criterion.}
    \label{fig:ke_etm2nc}
\vskip-5mm
\hspace*{-3mm} \includegraphics[width=15cm,height=61mm,angle=0]{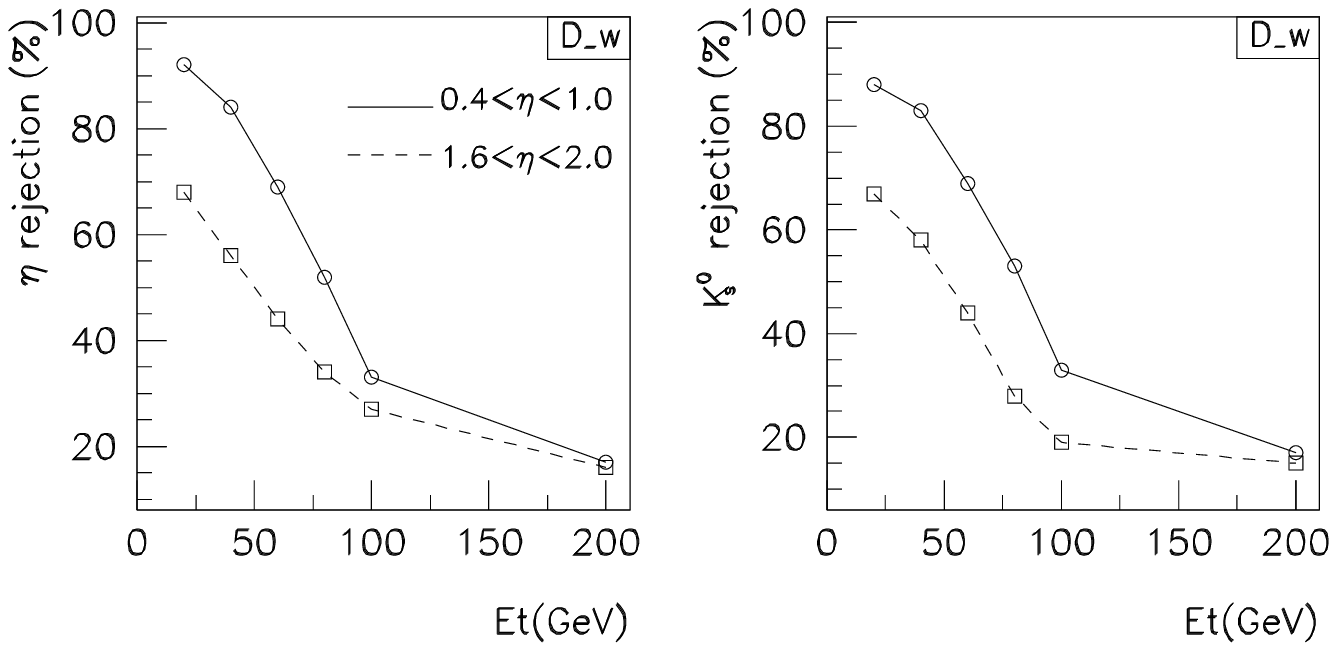}
\vskip-4mm
\hspace*{5mm} \footnotesize{Fig.~6: $\eta$ and $K^0_s$ rejection efficiencies for $\epsilon^{\gamma}_{eff}=90\%$.
Neutral decay channels only. $D_-w$ criterion.}
    \label{fig:ke_rw2nc}
  \end{figure}
\end{center}
~\\[-17mm]
the cuts chosen in \cite{PREP1} and \cite{PREP5}
decreases with growing $\Et$  faster than the ``$\gamma-brem$'' 
contribution. For this reason the bremsstrahlung photon background 
is more dangerous than the one from $\pi^0$ events.
It was also shown in \cite{PREP5} that the contribution to the total background from the
events containing the $\omega$ meson as a candidate for the direct photon is less than $1-2\%$.
Besides, the part of the neutral decay channels is only $8.5\%$ (Table 1). That is why
we have not considered in this section the rejection possibility of the $\omega$ mesons decaying via
neutral decay channels. 

\setcounter{figure}{6}


%
\section{Charged decay channels.}
%

Now let us return to  second group of the mesons
 decay channels that have charged daughter particles in the final state
(28$\%$ of $\eta$, 68.6$\%$ of $K^0_s$ and 88.8$\%$ of $\omega$ meson decays; see Table 1).

\begin{figure}
mesons having $\Et=40 ~GeV$ start at about $17-18 ~GeV$ for $\pi^\pm$ from $K^0_s$ decays
and at about $5-6~ GeV$ for $\pi^\pm$ produced in decays of $\eta$ and $\omega$ mesons.\\
\hspace*{9mm} So far we were discussing the spectra obtained at the level of PYTHIA simulation 
without the account
of detector effects. In Fig.~\ref{fig:pion5} we present two spectra obtained after CMSIM simulation 
of the calorimeter response to the propagation of charged pions  in the CMS detector
(the complete CMS setup was used). 
They correspond to the case of charged pions having 
$\Et=5~GeV$ and pseudorapidity $\eta$ equal 
to 0.4 and 1.7. The spectra are normalized to the total number of events and describe the 
distribution of transverse energy deposited in the HCAL (``$Et_-dep$''). They are built
using 2000 CMSIM simulated events with a single charged pion for each $\eta$ direction.
\vskip-10mm
\hspace*{-7mm} \includegraphics[width=15cm,height=13cm,angle=0]{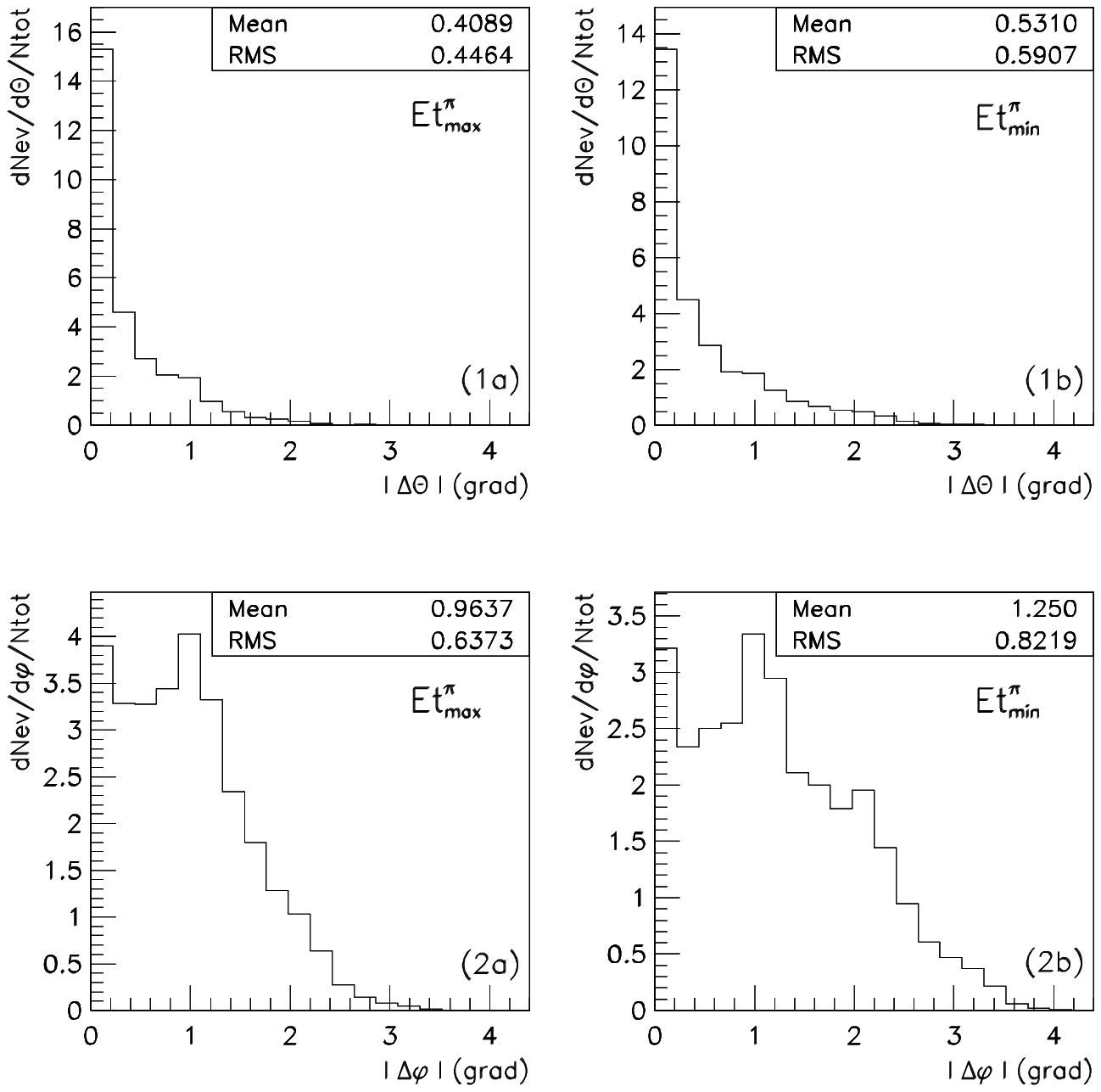}
\vskip-9mm
    \caption{
Absolute values of the difference in the $\theta$ and $\phi$ angles between $\pi^\pm$ and
photons or $\pi^0$ originated from the $\omega\to\pi^+\pi^-\pi^0$ decay. Plots {\sl 1a (1b)} and {\sl 2a (2b)}
correspond to the spectra over the angles between $\pi^0$ and the charged pion with the maximal(minimal)
$\Et^\pi$ in the decay channel.}
    \label{fig:omega_ang}
  \end{figure}

\begin{figure}
\hspace*{9mm} One can see from  Fig.~\ref{fig:pion5} that even a single  charged pion 
with $\Et=5~GeV$ has enough deposited energy to produce a noticeable signal in the HCAL in more 
than $90\%$ of events. 
In decays mentioned in Table 1 the charged mesons are produced in pairs. The $\Et$ spectra of 
the meson having the largest $\Et$ in this pair start at $\Et_{max}^\pi\geq 5~GeV$ as may be seen
from Fig.~\ref{fig:pt_eko}. So, one can expect that at least $90\%$ of this background may by rejected
by measuring the HCAL signal.\\
\hspace*{9mm} As we noted above, in reality we shall have a combined contribution of two charged pions
to the $5\times 5$ ECAL crystal cell window in one decay (see Figs.~\ref{fig:omega_ang} and 
\ref{fig:eta_ang})
and, thus, to HCAL towers behind it. The following CMSIM simulations were carried out to investigate this
contribution. First we  considered the $\eta$ meson decay 
\vskip-10mm
\hspace*{-7mm} \includegraphics[width=15cm,height=13cm,angle=0]{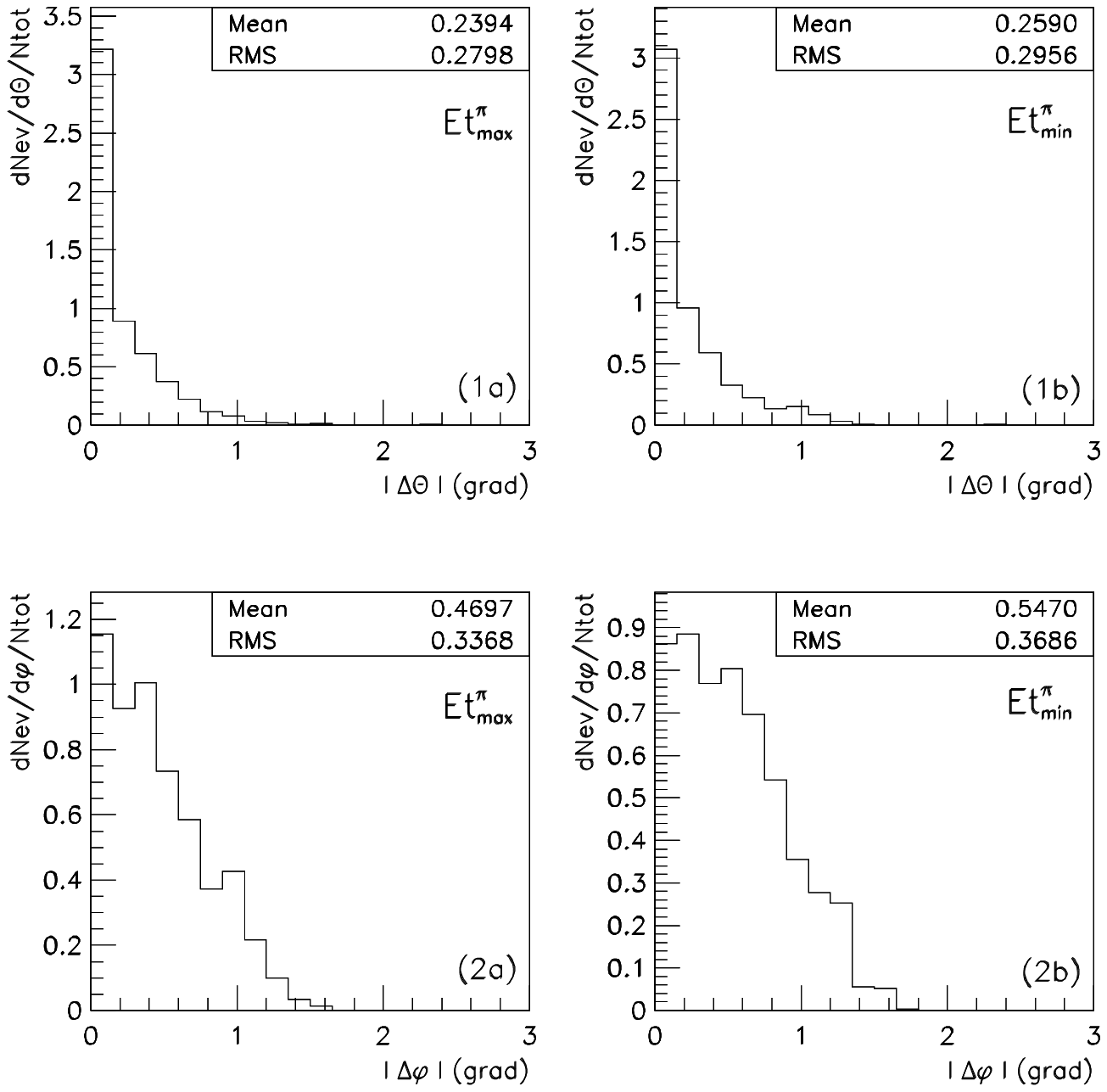}
\vskip-9mm
 \caption{
Absolute values of the difference in the $\theta$ and $\phi$ angles between charged
$\pi^\pm$ and $\gamma$ originated from the $\eta\to\pi^+\pi^-\gamma$ decay as well as 
between $\pi^\pm$ and $\pi^0$ from the $\eta\to\pi^+\pi^-\pi^0$ decay.
Plots {\sl 1a (1b)} and {\sl 2a (2b)} correspond to the spectra
over the angles between $\pi^0$ or $\gamma$ and charged pion with maximal (minimal) $\Et^\pi$
in the decay channel.}
    \label{fig:eta_ang}
  \end{figure}

We have  studied the angle separation between the charged pions originating
from $\omega$ and $\eta$ mesons and the neutral pion (as in a case
$\eta,\omega\to\pi^+\pi^-\pi^0)$ or photon ($\eta\to\pi^+\pi^-\gamma)$
produced in the same meson decay. For this aim the simulation of 1 million $pp$ events 
at $\sqrt{s}=14~TeV$ was carried out using PYTHIA 5.7 with the set of all QCD and SM subprocesses
having big cross sections. The minimal $\Pt$ of the  hard $2\to 2$ subprocess,
i.e. $\pth\equiv CKIN(3)$ parameter in PYTHIA, was taken to be $\pth=40~ GeV/c$.
The corresponding spectra normalized to unity are presented in Figs.~\ref{fig:omega_ang} and
\ref{fig:eta_ang} for $\Et^{\omega,\eta}\geq 30~ GeV$ separately for the charged pions with the maximal
and minimal $\Et$.
One can see (Fig.~\ref{fig:omega_ang}) that the charged pion 
deflects from the $\pi^0$ direction in the hadronic decay of $\omega$
by the angle $\Delta \theta= 0.4-0.5^\circ$  and by the angle 
$\Delta \phi=1.0-1.3^\circ$, on the average (the $\phi$ size of one crystal cell is about $\Delta\phi=1^\circ$).
The corresponding averaged deflections in the abovementioned  hadronic decays of the $\eta$ meson
are $\Delta\theta= 0.2^\circ$  and $\Delta\phi=0.5^\circ$ (see Fig.~\ref{fig:eta_ang}).
Thus, from the distributions shown in Figs.~\ref{fig:omega_ang} and \ref{fig:eta_ang} we may
conclude that practically in all events the charged pions 
enter the same ~$5\times 5$ ECAL crystal window as $\pi^0(\gamma)$ does and they may 
partially deposit their energy in the HCAL towers behind the ECAL cells that register the
$\pi^0(\gamma)$ signal
\footnote{Strictly speaking, these results are valid only
on the PYTHIA level of simulation and may be slightly modified when  the magnetic
field effect is taken into account in full GEANT simulation with CMSIM.}. 
The appearance of the corresponding signal in HCAL (see below for details of the CMSIM simulation) 
may allow, in principle, rejection of the events with the hadronic background.

\begin{figure}
\vskip-25mm
\hspace*{-5mm} \includegraphics[width=15cm,height=12cm,angle=0]{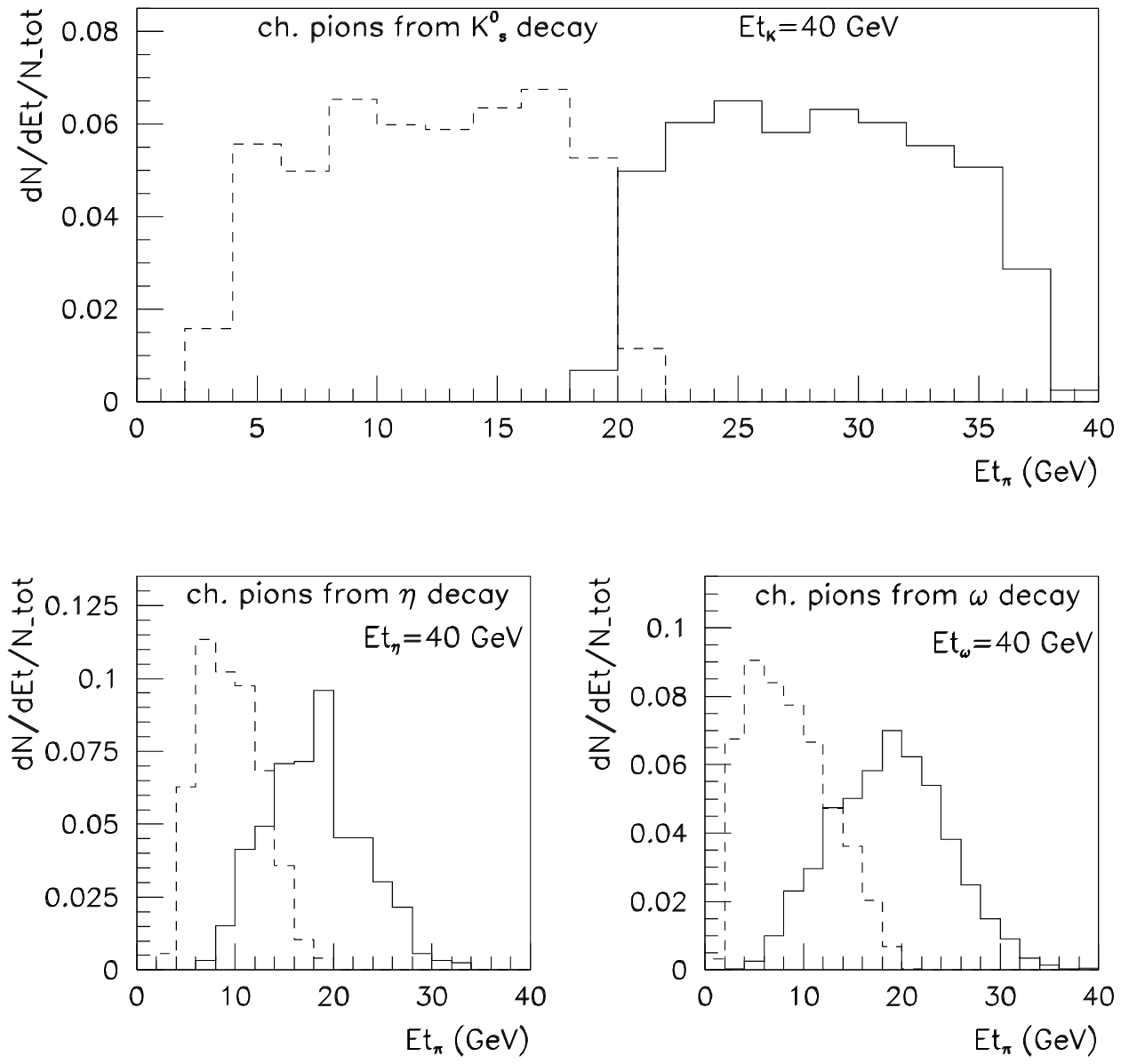}
\vskip-9mm
    \caption{$\Et$ spectra of charged pions from $K^0_s$, $\eta$ and $\omega$ meson decays.
Dashed and solid lines correspond to $\Et_{min}^{\pi}$ and
$\Et_{max}^{\pi}$ distributions respectively.}
    \label{fig:pt_eko}
\vskip-4mm
\hspace*{-5mm} \includegraphics[width=15cm,height=7cm,angle=0]{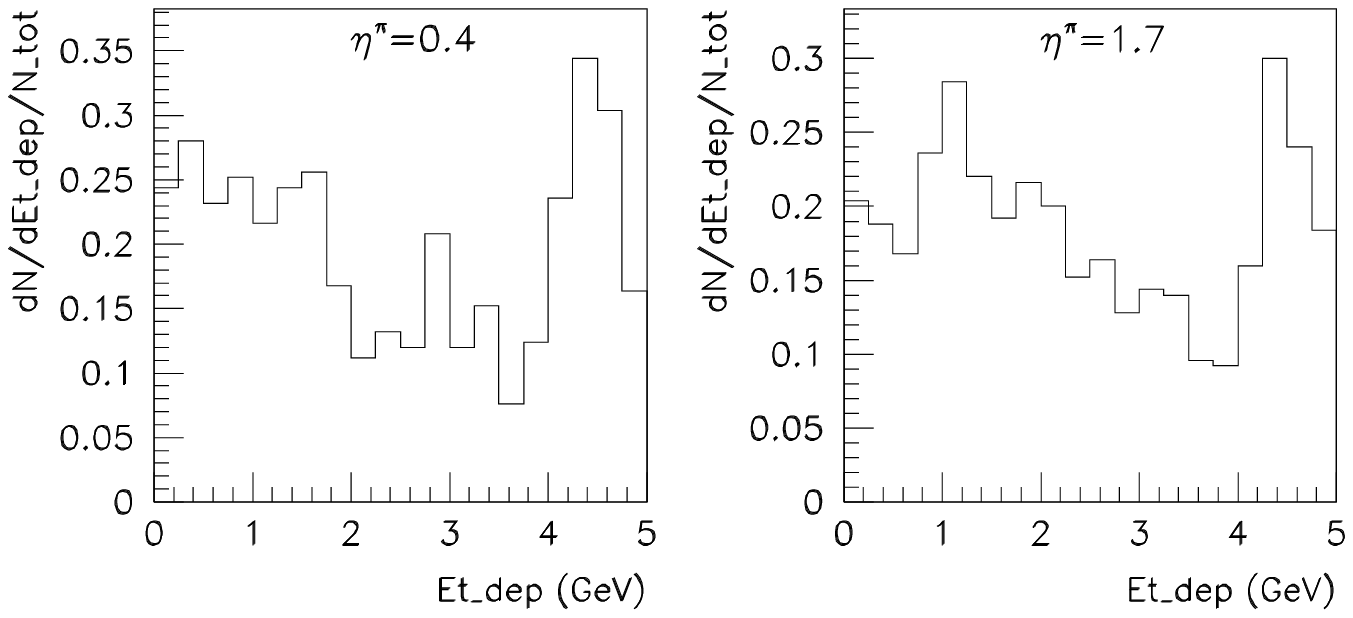}
{\vskip-9mm
    \caption{$\Et$ deposited in the HCAL by charged pions with initial
$\Et=5~GeV$. The left-hand plot corresponds to the Barrel region, the right-hand one corresponds
to the Endcap region.}
 \label{fig:pion5}
}
\end{figure}

The $\Et$ spectra of the charged pions originally produced in these channels were also studied
using PYTHIA. They are presented in Fig.~\ref{fig:pt_eko} for parent $\eta$, $K^0_s$ and $\omega$ mesons 
having $\Et=40 ~GeV$ separately for charged pions with  maximal $\Et^\pi_{max}$
and minimal $\Et^\pi_{min}$ in the decay event. We see that the $\Et^\pi_{max}$ distribution spectra for


\newpage
\noindent
as a typical example. 
To examine in what way the charged mode of  $\eta$ meson decay can fake the signal photon 
we forced $\eta$ meson  to decay only through the charged channels ($\eta\to\pi^+\pi^-\pi^0,
~\pi^+\pi^-\gamma$) with the pseudorapidities $\eta=0.4$ (Barrel region) and $\eta=1.7$ (Endcap region).
Three initial $\eta$ meson $\Et$ ranges  were chosen (for  both  
pseudorapidity values): $\Et^{\eta}_{init}$= $40\div 60$, $60\div 80$ and  $80\div 100~ GeV$. 
About $5000-7000$ single $\eta$ meson events for each $\eta$ direction and $\Et$ 
interval were generated for this aim (again with the complete setup of the CMS detector).

Since we want here to separate the $\eta$ meson background one the basis of the presence of
its hadronic  decay products around a photon or a neutral pion (see Table 1), 
we apply  here the isolation criteria formulated in \cite{PREP1}.
It turned out that in the abovementioned generated samples of $\eta$ meson events 
the number of events with the 
transverse energy inside the $5\times 5$ ECAL crystal cell window 
$\Et^{5\times 5}_{ECAL}\geq 40 ~GeV$ is not quite sufficient for our analysis. 
As a good approximation to the  value declared in \cite{PREP1} ($\Ptg\gt40 ~GeV/c$)
we choose here $\Et^{5\times 5}_{ECAL}\geq 35 ~GeV$ as a lower cut.

\begin{figure}
\vskip-20mm
\hspace*{-7mm} \includegraphics[width=15cm,height=17.5cm,angle=0]{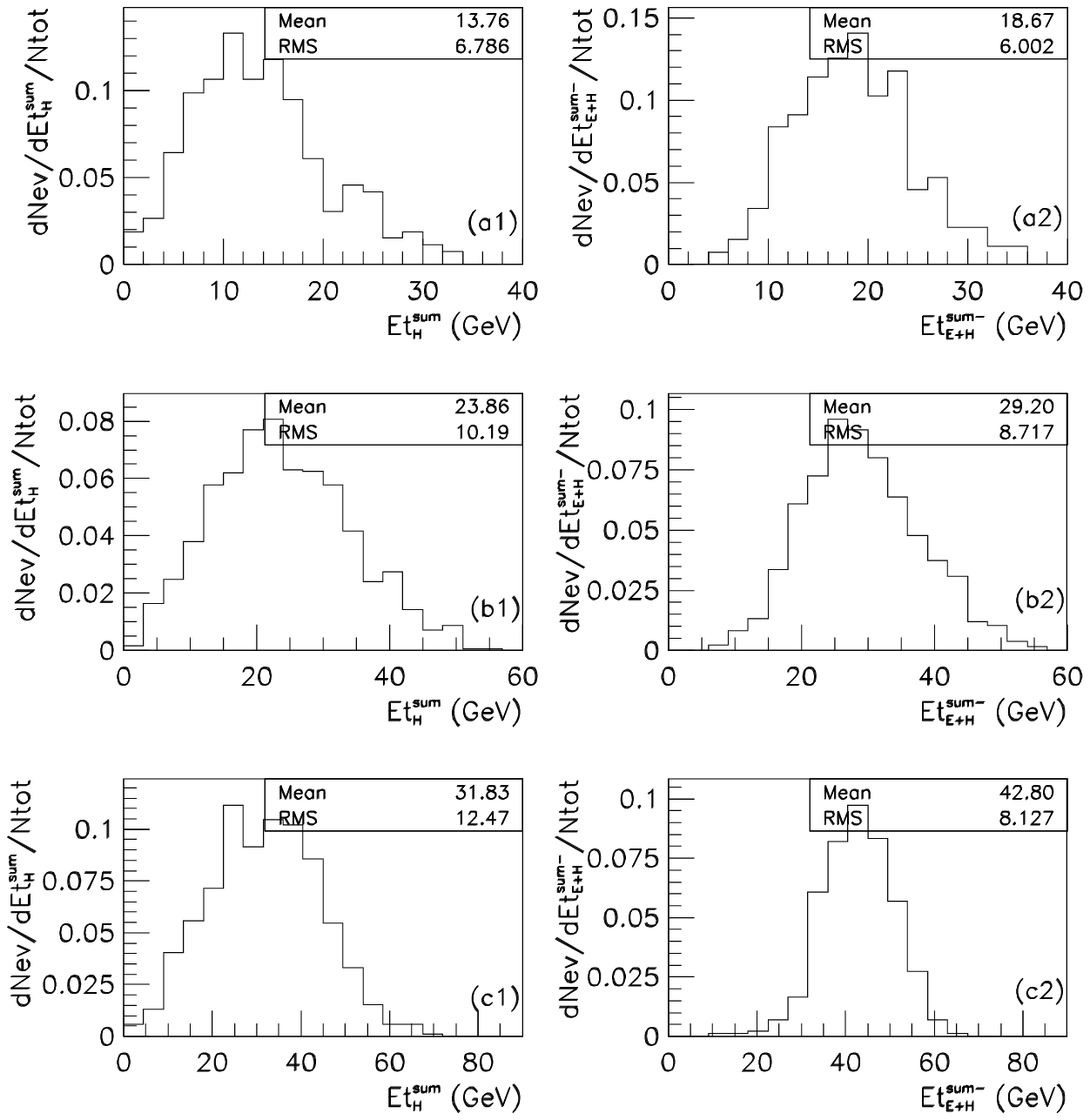}
\vskip-7mm
 \caption{Normalized distributions of number of the $\eta$ meson charged decay events over
$\Et$ deposited within the radius $R(\eta,\phi)=0.7$, counted from the most energetic ECAL crystal cell, 
(1) in HCAL (a1, b1, c1) and (2) in ``ECAL+HCAL'' cells beyond the
ECAL $5\times 5$ crystal cell window (a2, b2, c2). The 1st row (a1, a2) corresponds to the initial 
$\Et$ of the $\eta$ meson in the
range $40\!\leq\!\Et^{\eta}_{init}\!\leq\! 60 ~GeV$, the 2nd row to that in the range
$60\!\leq\!\Et^{\eta}_{init}\!\leq\! 80 ~GeV$ and the 3rd to that in the range
$80\!\leq\!\Et^{\eta}_{init}\!\leq\! 100 ~GeV$.
All distributions were obtained with the cut $\Et^{5\times 5}_{ECAL}\!\!\geq\!\!35 ~GeV$.
The Barrel case ($\eta=0.4$).}  
    \label{fig:eta_bar}
  \end{figure}

\begin{figure}
\vskip-20mm
\hspace*{-7mm} \includegraphics[width=15cm,height=17.5cm,angle=0]{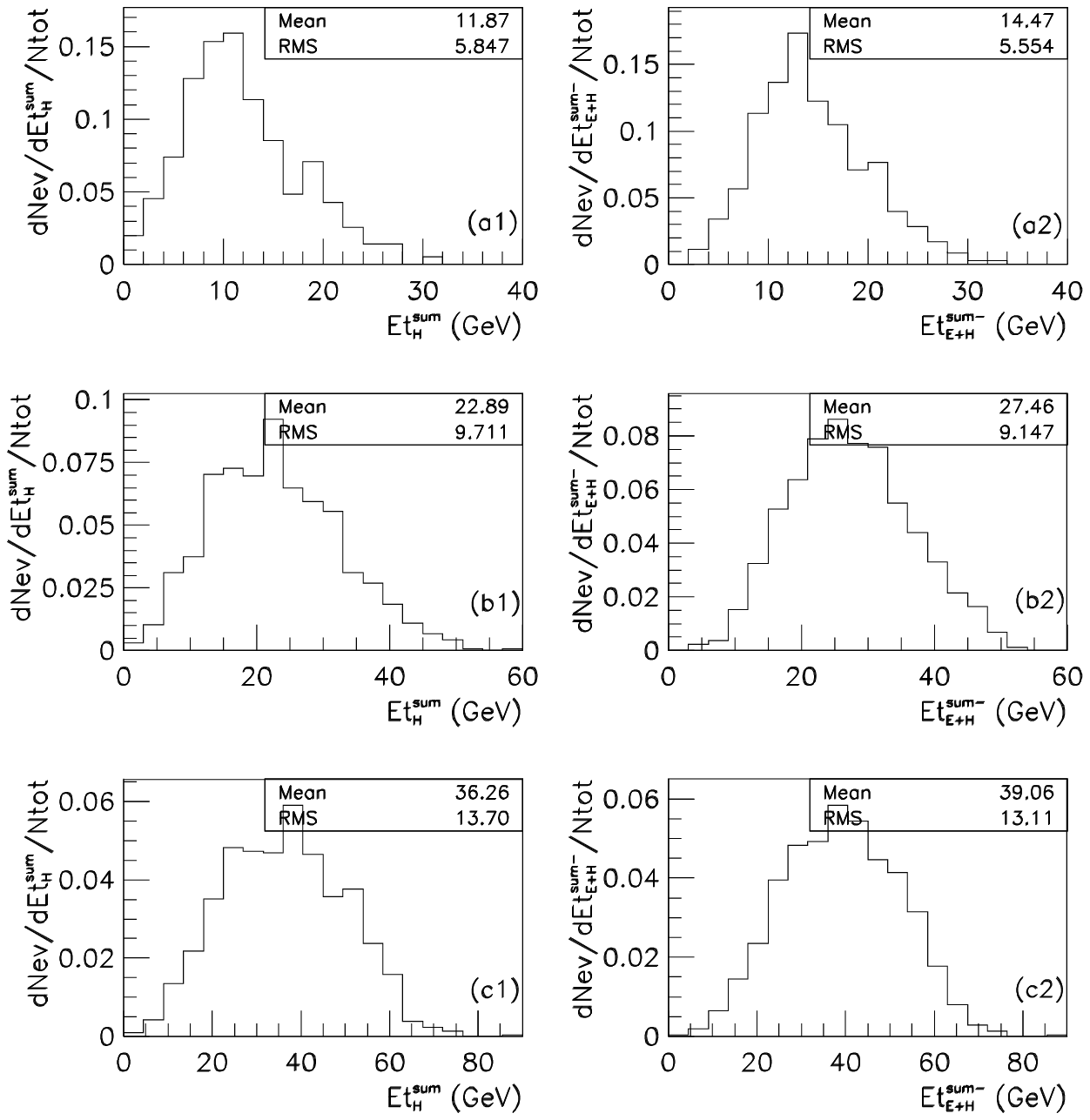}
\vskip-7mm
 \caption{Normalized distributions of number of the $\eta$ meson charged decay events over
$\Et$ deposited within the radius $R(\eta,\phi)=0.7$, counted from the most energetic ECAL crystal cell, 
(1) in HCAL (a1, b1, c1) and (2) in ``ECAL+HCAL'' cells beyond the
ECAL $5\times 5$ crystal cell window (a2, b2, c2). The 1st row (a1, a2) corresponds to the initial 
$\Et$ of the $\eta$ meson in the
range $40\!\leq\!\Et^{\eta}_{init}\!\leq\! 60 ~GeV$, the 2nd row to that in the range
$60\!\leq\!\Et^{\eta}_{init}\!\leq\! 80 ~GeV$ and the 3rd to that in the range
$80\!\leq\!\Et^{\eta}_{init}\!\leq\! 100 ~GeV$.
All distributions were obtained with the cut $\Et^{5\times 5}_{ECAL}\!\!\geq\!\!35 ~GeV$.
The Endcap case ($\eta=1.7$).}     
\label{fig:eta_end}
\end{figure}

The spectra of the hadronic transverse energy deposited in HCAL towers behind the $5\times 5$ ECAL
crystal cell window 
\footnote{containing $\Et^{5\times 5}_{ECAL}\geq 35 ~GeV$}
$\Et_{H}^{sum}$ are given in plots (a1), (b1) and (c1) of 
Figs.~\ref{fig:eta_bar} and \ref{fig:eta_end} for the Barrel and the Endcap, respectively,
for three abovementioned $\Et$ ranges.
In these plots $\Et_{H}^{sum}$ is defined as a sum of the transverse energies $\Et^i$\\[-9pt]
\begin{equation}
\Et^{sum}_{H}=\sum_{i}\Et^i 
\label{eq:eths}
\end{equation}
deposited in each $i$th HCAL tower fired by $\pi^\pm$ showers. 
No events were found with $\Et$ deposited in the HCAL with 
$\Et_{H}^{sum} \leq 0.5 ~GeV$. In the intervals  $\Et^{\eta}_{init}=40\div 60$ and
$60\div 80 ~GeV$ only about $0.1-0.2 \%$ of the  
events passed the cut $\Et^{5\times 5}_{ECAL}\!\!>\!\!35 ~GeV$ have $\Et_{H}^{sum} \leq 1 ~GeV$.
No events with $Et_{H}^{sum}\leq 1 ~GeV$ were found
for $\Et^{\eta}_{init}=80\div 100~GeV$ for both values $\eta=0.4$ and $1.7$.

\begin{figure}
\vskip-15mm
\hspace*{-7mm} \includegraphics[width=15cm,height=18cm,angle=0]{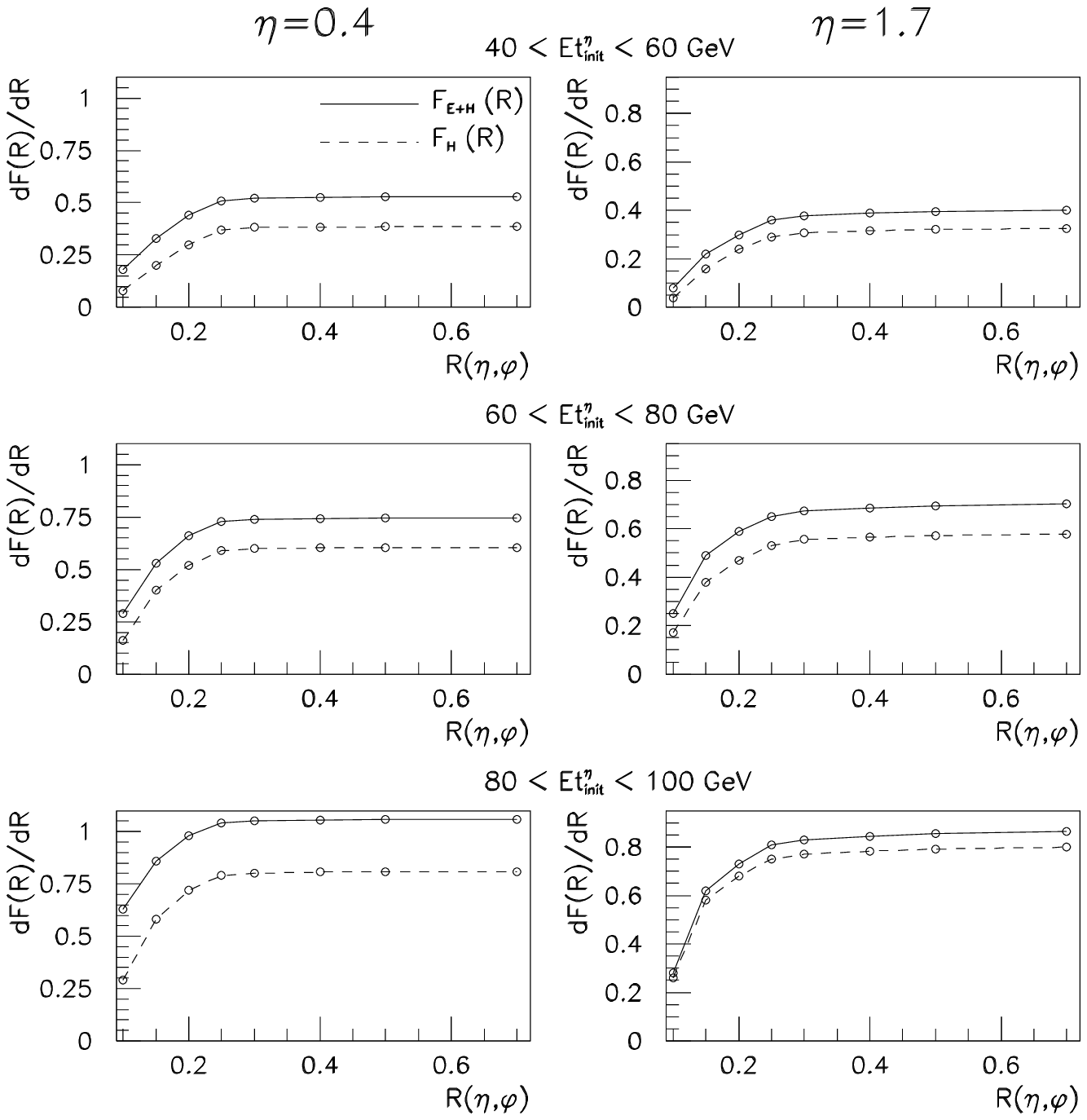}
\vskip-7mm
 \caption{Distributions of the ratios of the $\Et$ deposited in HCAL (dashed line) and in 
``ECAL+HCAL'' cells  beyond the ECAL $5\times 5$ crystal cell window (solid line) to the $\Et$ 
deposited inside this window
as the function of the distance from the center of gravity $R(\eta,\phi)$,
counted from the most energetic ECAL crystal cell. Left-hand and right-hand columns correspond to
the Barrel ($\eta=0.4$) and Endcap ($\eta=1.6$) cases.}     
\label{fig:et_r_ht}
\end{figure}

Let us remind that in \cite{PREP1} we have chosen  the value of
the isolation cone radius around a $\gamma^{dir}$-candidate to be $R^{\gamma}_{isol}=0.7$. 
It is useful to find out what radius of  the hadronic energy shower from the $\eta$ meson decay
may be in reality. For this aim we calculate 
$F_H(R)\equiv\Et^{sum}_{H}(R)/\Et^{5\times 5}_{ECAL}$.
The value of $\Et^{sum}_{H}(R)$ differs from $\Et^{sum}_{H}$ defined by (\ref{eq:eths}) by including 
to the sum only those towers that fit into the circle of some radius $R(\eta,\phi)$
counted from the most energetic ECAL crystal cell of the abovedescribed  
$5\times 5$ ECAL crystal cell window, i.e.\\[-8pt]
\begin{equation}
\Et^{sum}_{H}(R)=\sum_{i\in R}\Et^i.
\label{eq:eths_r}
\end{equation}
The dependence of $F_H$ on this radius $R$
describes the $\Et$ saturation of the space around the most energetic ECAL cell
in the $5\times 5$ ECAL crystal window. This dependence is shown in Fig.~\ref{fig:et_r_ht} by 
the dashed line for three ranges of the initial transverse energies and two values of $\eta$.
The quantity reaches the saturation values in the range of $R(\eta,\phi)\approx 0.3-0.4$
that has a meaning of a real radius of the hadronic energy deposition in the HCAL for the 
single $\eta$ meson decay. This value agrees with the value $R^{\gamma}_{isol}=0.7$ chosen 
in \cite{PREP1}.

Now let us see what size of an additional $\Et$ would be added to the isolation cone
by the energy of charged pions deposited in ECAL cells surrounding the $5\times 5$ ECAL crystal window
(containing $\Et^{5\times 5}_{ECAL}\geq 35 ~GeV$). 
Let us define the value of $\Et^{sum-}_{E+H}$ as a scalar sum of $\Et$ deposited inside
the calorimeter ECAL+HCAL cells around the ECAL $5\times 5$ crystal cell window
(i.e.  with subtraction of $\Et$ deposited inside the ECAL $5\times 5$ crystal cell 
window itself):\\[-8pt]
\begin{equation}
\Et^{sum-}_{E+H}=\!\!\!\!\!\sum_{i\in ECAL+HCAL}\!\!\!\!\!\!\!\!\!\!
\Et^i-\Et^{5\times 5}_{ECAL}. 
\label{eq:etehs}
\end{equation}

Plots (a2), (b2) and (c2) in Figs.~\ref{fig:eta_bar} and \ref{fig:eta_end}
include the spectra of $\Et$ deposited in the ECAL+HCAL calorimeter cells that are
beyond the $5\times 5$ ECAL crystal cell window containing the most energetic crystal cell at
its center, but within the radius  $R(\eta,\phi)=0.7$ counted from the center of this cell. 
They are given again for three different ranges of
$\Et^{\eta}_{init}$ and two values of pseudorapity $\eta=0.4$ and $1.7$.
We see that all spectra  in Figs.~\ref{fig:eta_bar} and \ref{fig:eta_end}
steeply go to zero  in the region of small $\Et$ values.
It allows the background from the charged $\eta$ meson decays to be reduced
by limiting $\Et$ deposited in the ECAL+HCAL cells within the radius $R(\eta,\phi)=0.7$.
One can see from the right-hand columns of Figs.~\ref{fig:eta_bar} and \ref{fig:eta_end}
that there are no events with $\Et$ 
less than $2 ~GeV$
for the entire range of $\Et^{\eta}_{init}$ $40\div 100 ~GeV$ and both
pseudorapidity values. This fact partially explains our choice of the direct photon isolation cut 
in \cite{PREP1} with $\Et^{isol}\!\leq 2~ GeV/c$.

Let us introduce another variable $F_{E+H}(R)$
which has a meaning of the ratio of $\Et^{sum-}_{E+H}(R)$, i.e. the value of 
$\Et^{sum-}_{E+H}$ taken in analogy with (\ref{eq:eths_r}) for $i\in R$, 
to $\Et^{5\times 5}_{ECAL}$:\\[-7pt]
\begin{equation}
F_{E+H}(R) = \Et^{sum-}_{E+H}(R)/\Et^{5\times 5}_{ECAL}.
\label{eq:f_er_eh}
\end{equation}
Its dependence on the distance from the center of gravity $R(\eta,\phi)$
(counted from the most energetic ECAL crystal cell) is shown in Fig.~\ref{fig:et_r_ht} by solid lines.
This variable also reaches its saturation in the same range of $R(\eta,\phi)\approx 0.3-0.4$.
The difference between solid and dashed curves defines the percentage of transverse energy deposited 
by electromagnetic showers produced by charged pions from the
$\eta$ meson decay in ECAL cells around the $5\times 5$ ECAL crystal window containing
the $\gamma$-candidate with $\Et^{5\times 5}_{ECAL}\geq 35 ~GeV$. This difference
 does not exceed $14-20\%$ in the Barrel and $7-9\%$ in the Endcap.
Its dependence on the distance in the region of small $R(\eta,\phi)$ is evident from these pictures.  

Multiplying the ratios $F_H(R)$ and $F_{E+H}(R)$ in the region where they reach their saturation 
by $\Et^{5\times 5}_{ECAL}$,
 one can obtain mean $\Et$ of the corresponding distributions in the plots of 
Figs.~\ref{fig:eta_bar} and \ref{fig:eta_end} (with the RMS values shown in the plots).
 
So far we were discussing the charged decay channels of the $\eta$ meson alone.
As for the $\omega\to\pi^+\pi^-\pi^0$ and $K^0_s\to\pi^+\pi^-$ decays, let us note that the former
is analogous to the $\eta\to\pi^+\pi^-\pi^0$ decay (even a bit easier from the viewpoint of
rejection because the value of the charged pion deflection by angles $\theta$ and $\phi$ 
 from the $\pi^0$ direction is, on the average, twice larger).
The channel $K^0_s\to\pi^+\pi^-$ does not have a neutral particle. In this case
the probability that no signal in the HCAL will be observed after imposing
the cut $\Et^{5\times 5}_{ECAL}\!\!\geq\!\! 35 ~GeV$ is very small
\footnote{and it is defined by the probability of the case when both charged pions deposit their energies
in the ECAL alone}.

Thus, we can conclude that to suppress contributions from the meson decays via charged channels one can
impose the upper cuts \\[-5mm]
\begin{itemize}
\item  on $\Et$ deposited only in the HCAL in the isolation cone of radius 
$R^{\gamma}_{isol}=0.3$ for the Barrel and for the Endcap
(according to the dashed lines in Fig.~\ref{fig:et_r_ht}).
We can require that $\Et^{sum}_{H}\!\leq\! \Et^{thr}_{H}$.
The threshold $\Et^{thr}_{H}$ depends on the parent meson $\Et$.
We choose this cut to be equal to $0.5-1~GeV$ for $\Et^{\gamma}=40 ~GeV$,
gradually increasing it to $\Et^{thr}_{H}\!\!=\!\!2~GeV$ for $\Et^{\gamma}=200~GeV$
(see left-hand plots of  Figs.~\ref{fig:eta_bar} and \ref{fig:eta_end}).\\[-4mm]
\item on  $\Et$ deposited in the calorimeter ``ECAL+HCAL'' cells
within the radius $R^{\gamma}_{isol}=0.3$ beyond the $5\times 5$ ECAL window. We can require that
~$\Et^{sum-}_{E+H}\!\!\leq\!\!\Et^{thr}_{E+H}$ (with $\Et^{thr}_{E+H}=2-5~ GeV/c$)
in accordance with the right-hand plots of Figs.~\ref{fig:eta_bar} and \ref{fig:eta_end}.
These values justify our choice of the absolute isolation cut in \cite{PREP1}
\footnote{About $65-70 \%$ of the signal events with the direct photon satisfy this requirement
(see \cite{PREP5}).}.
\end{itemize}
~\\[-10mm]

Therefore, we see that the two criteria introduced above may allow the charged hadronic decay channels of
$\eta,\omega$ and $K^0_s$ mesons to be suppressed with a good efficiency
(about $98\%$, depending on the exact value of $\Et^{thr}_{E+H}$; see 
Figs.~\ref{fig:eta_bar} and \ref{fig:eta_end}).

We have considered the extreme case of the single meson decaying through charged channels.
Certainly, those criteria would be much more efficient 
in the case of real $pp$ collisions where
$\eta,\omega$ and $K^0_s$ mesons may have a hadronic accompaniment in the isolation cone around them.

To conclude, let us add that in selection of
the \gpj events in real $pp$ collisions these criteria can be strengthened
by additionally requiring that there should be no of 
a charged particle track with $\Et\gt1~GeV$  within the isolation cone of
 radius $R^{\gamma}_{isol}\!=\!0.7$ around a photon candidate in the opposite (in 
$\phi$)
direction to a jet (see also \cite{PREP1}, \cite{PREP5}).
Additional rejection of the charged pions that do not reveal themselves in the HCAL below 
some threshold depends mainly on the track finding efficiency for them.


\end{document}